\documentclass[12pt,a4paper,dvipdfm]{article}
\usepackage{graphicx}

\usepackage[ruled,lined,algonl,boxed]{algorithm2e}

\begin{document}
\newcommand{\up}{\vspace*{-0.05cm}}
\newcommand{\qed}{\hfill$\rule{.05in}{.1in}$\vspace{.3cm}}
\newcommand{\pf}{\noindent{\bf Proof: }}
\newtheorem{thm}{Theorem}
\newtheorem{lem}{Lemma}
\newtheorem{prop}{Proposition}
\newtheorem{prob}{Problem}
\newtheorem{ex}{Example}
\newtheorem{cor}{Corollary}
\newtheorem{conj}{Conjecture}
\newtheorem{cl}{Claim}
\newtheorem{df}{Definition}
\newtheorem{rem}{Remark}
\newcommand{\beq}{\begin{equation}}
\newcommand{\eeq}{\end{equation}}
\newcommand{\<}[1]{\left\langle{#1}\right\rangle}
\newcommand{\be}{\beta}
\newcommand{\ee}{\end{enumerate}}
\newcommand{\Bul}{\mbox{$\bullet$ } }
\newcommand{\al}{\alpha}
\newcommand{\ep}{\epsilon}
\newcommand{\si}{\sigma}
\newcommand{\om}{\omega}
\newcommand{\la}{\lambda}
\newcommand{\La}{\Lambda}
\newcommand{\Ga}{\Gamma}
\newcommand{\ga}{\gamma}
\newcommand{\im}{\Rightarrow}
\newcommand{\2}{\vspace{.2cm}}
\newcommand{\es}{\emptyset}

\title{\huge\bf The probabilistic approach\\ to limited packings in graphs}
\author{Andrei Gagarin\footnote{e-mail: {\tt andrei.gagarin@rhul.ac.uk}}\\ 
{\footnotesize Department of Computer Science, Royal Holloway, University of London}\\
{\footnotesize Egham, Surrey, TW20 0EX, UK}\vspace{4mm}\\
Vadim Zverovich\footnote{e-mail: {\tt vadim.zverovich@uwe.ac.uk}}\\
{\footnotesize Department of Mathematics and Statistics, University of the West of England}\\
{\footnotesize Bristol, BS16 1QY, UK}
\\
}
\maketitle


\maketitle

\begin{abstract}
We consider (closed neighbourhood) packings and their generalization in graphs.
A vertex set $X$ in a graph $G$ is a {\it $k$-limited packing} if for any vertex $v\in V(G)$,
$
\left|N[v] \cap X\right| \le k,
$ 
where $N[v]$ is the closed neighbourhood of $v$. 
The {\it $k$-limited packing number} $L_k(G)$ of a graph $G$ is the largest size of 
a $k$-limited packing in $G$. Limited packing problems can be considered as secure facility location problems in networks.

In this paper, we develop a new probabilistic approach to limited packings in graphs, resulting in lower bounds for the $k$-limited packing number and a randomized algorithm to find $k$-limited packings satisfying the bounds. In particular, we prove that for any graph $G$ of order $n$ with maximum vertex degree $\Delta$,
$$
L_k(G) \ge {k n \over (k+1) \sqrt[k]{\pmatrix{\Delta \cr k} (\Delta +1)} }.
$$
The problem of finding a maximum size $k$-limited packing is known to be $NP$-complete even in split or bipartite graphs.

\bigskip


\end{abstract}

\bigskip\bigskip
\section{Introduction}
\label{intro}
\noindent We consider simple undirected graphs. If not specified otherwise, standard graph-theoretic terminology and notations are used (e.g., see \cite{AS00,B79}). We are interested in the classical packings and packing numbers of graphs as introduced in \cite{MM75}, and their generalization, called limited packings and limited packing numbers, respectively, as presented in \cite{GGHR10}. In the literature, the classical packings are often referred to under different names: for example, as (distance) $2$-packings \cite{MM75,TV91}, closed neighborhood packings \cite{RSS06} or strong stable sets \cite{HS85}. They can also be considered as generalizations of independent (stable) sets which, following the terminology of \cite{MM75}, would be (distance) $1$-packings.

Formally, a vertex set $X$ in a graph $G$ is a {\it $k$-limited packing} if for any vertex $v\in V(G)$,
$$
\left|N[v] \cap X\right| \le k,
$$ 
where $N[v]$ is the closed neighbourhood of $v$. 
The {\it $k$-limited packing number} $L_k(G)$ of a graph $G$ is the maximum size of 
a $k$-limited packing in $G$. In these terms, the classical (distance) $2$-packings are $1$-limited packings, and hence $\rho(G)=L_1(G)$, where $\rho(G)$ is the 2-packing number.

The problem of finding a $2$-packing ($1$-limited packing) of maximum size is shown to be $NP$-complete by Hochbaum and Schmoys \cite{HS85}. In \cite{DLN11}, it is shown that the problem of finding a maximum size $k$-limited packing is $NP$-complete even for the classes of split and bipartite graphs.

Graphs usually serve as underlying models for networks. A number of interesting application scenarios of limited packings are described in \cite{GGHR10}, including network security, market saturation, and codes. These and others can be summarized as secure location or distribution of facilities in a network. In a more general sense, these problems can be viewed as (maximization) facility location problems to place/distribute in a given network as many resources as possible subject to some (security) constraints.

$2$-Packings ($1$-limited packings) are well-studied in the literature from the structural and algorithmic point of view (e.g., see \cite{HS85, MM75, RSS06, S12}) and in connection with other graph parameters (e.g., see \cite{BHV09, HLR11, MM75, RSS06, TV91}). In particular, several papers discuss connections between packings and dominating sets in graphs (e.g., see \cite{BHV09, DLN11, GGHR10, HLR11, RSS06}). Although the formal definitions for packings and dominating sets may appear to be similar, the problems have a very different nature: one of the problems is a maximization problem not to break some (security) constraints, and the other is a minimization problem to satisfy some reliability requirements. For example, given a graph $G$, the definitions imply a simple inequality $\rho(G)\le\gamma(G)$, where $\gamma(G)$ is the domination number of $G$ (e.g., see \cite{RSS06}). However, the difference between $\rho(G)$ and $\gamma(G)$ can be arbitrarily large as illustrated in \cite{BHV09}: $\rho(K_n\times K_n)=1$ for the Cartesian product of complete graphs, but $\gamma(K_n\times K_n)=n$.

In this paper, we develop the probabilistic method for $k$-limited packings in general and for $2$-packings ($1$-limited packings) in particular. In Section 2 we present the probabilistic construction and use it to derive two lower bounds for the $k$-limited packing number $L_k(G)$. The construction implies a randomized algorithm to find $k$-limited packings satisfying the lower bounds. 
The algorithm and its analysis are presented in Section 3.
Section 4 shows that one of the lower bounds is asymptotically sharp. Finally, Section 5 provides upper bounds for $L_k(G)$, e.g. in terms of the $k$-tuple domination number $\gamma_{\times k}(G)$.

Notice that the probabilistic construction and approach are different from the probabilistic constructions used for independent sets and deriving the lower bounds for the independence number $\alpha(G)$ in \cite{AS00}, pp.\,27--28,\,91--92. In terms of packings, an independent set in a graph $G$ is a distance $1$-packing: for any two vertices in an independent set, the distance between them in $G$ is greater than $1$. To the best of our knowledge, the proposed probabilistic method is a new approach to work on packings and related maximization problems.

\section{The probabilistic construction and lower bounds}
\label{main}
Let $\Delta$ = $\Delta(G)$ denote the maximum vertex degree in a graph $G$. Notice that $L_k(G)=n$ when $k\ge\Delta+1$.
We define
$$
 {c}_t = {c}_t (G) =  \pmatrix{\Delta \cr t}
 \quad \mbox{and} \quad
 {\tilde c}_t = {\tilde c}_t (G) =  \pmatrix{\Delta+1 \cr t}.
$$
In what follows, we put ${a \choose b}=0$ if $b>a$.

The following theorem gives a new lower bound for the $k$-limited packing number. 
It may be pointed out that the probabilistic construction used in the
proof of Theorem \ref{th1}  implies a randomized algorithm for finding a $k$-limited packing  set, whose size satisfies the bound of Theorem \ref{th1} with a positive
probability (see Algorithm 1 in Section 3).

\begin{thm} \label{th1}
For any graph $G$ of order $n$ with $\Delta\ge k \ge 1$,
\begin{equation} \label{main_bound}
L_k(G) \ge {k n \over {\tilde c}_{k+1}
^{1/k} \; (1+k)^{1+1/k}}.
\end{equation}
\end{thm}

\begin{pf}
Let $A$ be a set formed by an
independent choice of vertices of $G$, where each vertex is
selected with the probability 
\begin{equation}
\label{probability}
p = \left({1 \over {\tilde c}_{k+1} \; (1+k)}\right)^{1/k}. 
\end{equation}
 
For
$m=k,...,\Delta$, we denote 
$$ 
A_m = \{v\in A : |N(v)\cap A|=m\}.
$$ 
For each set $A_m$, we form a set $A'_m$ in the following
way. For every vertex $v\in A_m$, we take $m-(k-1)$ neighbours from
$N(v) \cap A$ and add them to $A'_m$. Such neighbours always exist
because $m \ge k$. It is obvious that 
$$
|A'_m|\le (m-k+1)|A_m|.
$$ 

For $m=k+1,...,\Delta$, let us denote 
$$ 
B_m = \{v\in V(G)-A : |N(v)\cap A|=m\}.
$$
For each set $B_m$, we form a set $B'_m$ by taking
$m-k$ neighbours from $N(v) \cap A$ for every vertex $v\in B_m$. We
have $$|B'_m|\le (m-k)|B_m|.$$

Let us construct the set $X$ as follows: 
$$ 
X = A -
\left(\bigcup_{m=k}^{\Delta} A'_m\right) -
\left(\bigcup_{m=k+1}^{\Delta} B'_m\right).
$$
It is easy to see
that $X$ is a $k$-limited packing in $G$. The expectation of $|X|$
is
\begin{eqnarray*}
{\mathbf E}[|X|] &\ge& {\mathbf E}\left[|A| - \sum_{m=k}^{\Delta}|A'_m| - \sum_{m=k+1}^{\Delta}|B'_m|\right]\\
&\ge& {\mathbf E}\left[|A| - \sum_{m=k}^{\Delta}(m-k+1) |A_m|  - \sum_{m=k+1}^{\Delta}(m-k)|B_m|\right]\\
&=& pn - \sum_{m=k}^{\Delta}(m-k+1){\mathbf E}[|A_m|]  - \sum_{m=k+1}^{\Delta}(m-k){\mathbf E}[|B_m|].
\end{eqnarray*}

Let us denote the vertices of $G$ by $v_1, v_2,..., v_n$ and the corresponding vertex degrees by $d_1, d_2,..., d_n$.
We will need the following lemma:

\begin{lem}\label{lemma1}
If $p = \left({1 \over {\tilde c}_{k+1} \; (1+k)}\right)^{1/k}$, then, 
for any vertex $v_i\in V(G)$,
\begin{equation} \label{eq2}
\pmatrix{d_i \cr m} (1-p)^{d_i-m} \le \pmatrix{\Delta \cr m} (1-p)^{\Delta-m}.
\end{equation}
\end{lem}

\begin{pf}
The inequality (\ref{eq2})  holds if $d_i=\Delta$. It is also true if $d_i<m$ because in this case $\pmatrix{d_i \cr m}=0$.
Thus, we may assume that
$$
m\le d_i < \Delta.
$$
Now, it is easy to see that inequality (\ref{eq2}) is equivalent to the following:
\begin{equation} \label{eq3}
(1-p)^{\Delta-d_i} \ge \pmatrix{d_i \cr m} / \pmatrix{\Delta \cr m} = 
{(\Delta -m)!/(d_i-m)! \over \Delta!/d_i!}
=\prod_{i=0}^{\Delta -d_i -1}{\Delta -m-i \over \Delta -i}.
\end{equation} 

Further, 
$\Delta\ge k$ implies ${\Delta\over k} \le {\Delta-i\over k-i}$, where $0\le i\le k-1$. 
Taking into account that $\Delta >0$, we obtain
$$
\left({\Delta\over k} \right)^k \le \prod_{i=0}^{k-1}{\Delta-i \over k-i} = c_k < {\tilde c}_{k+1} (1+k)
$$
or
$$
{ 1 \over {\tilde c}_{k+1}\; (1+k)} < \left({k\over \Delta} \right)^k.
$$
Thus,
$$
p^k < \left({k\over \Delta} \right)^k  \quad  \mbox{or}  \quad p < {k\over \Delta} \le {m\over \Delta}. 
$$
We have $ p <   {m\over \Delta}$, which is equivalent to $1-p > {\Delta-m \over \Delta}$.
Therefore,
$$
(1-p)^{\Delta-d_i} > \left({\Delta - m \over \Delta}\right)^{\Delta-d_i} \ge  \prod_{i=0}^{\Delta -d_i -1}{\Delta -m-i \over \Delta -i},
$$
as required in (\ref{eq3}). 
\end{pf}
\qed

\bigskip
Now we go on with the proof of Theorem \ref{th1}. By Lemma \ref{lemma1},
\begin{eqnarray*}
{\mathbf E}[|A_m|]  &=&  \sum_{i=1}^n {\mathbf P}[v_i\in A_m]\\
   &=&  \sum_{i=1}^n p \pmatrix{d_i \cr m} p^m (1-p)^{d_i-m}\\
 &\le& p^{m+1} \sum_{i=1}^n \pmatrix{\Delta \cr m} (1-p)^{\Delta-m} \\
 &=& p^{m+1} (1-p)^{\Delta-m} {c}_m n
\end{eqnarray*}
and
\begin{eqnarray*}
{\mathbf E}[|B_m|]  &=&  \sum_{i=1}^n {\mathbf P}[v_i\in B_m]\\
   &=&  \sum_{i=1}^n (1-p) \pmatrix{d_i \cr m} p^m (1-p)^{d_i-m}\\
 &\le& p^{m} \sum_{i=1}^n \pmatrix{\Delta \cr m} (1-p)^{\Delta-m+1} \\
 &=& p^{m} (1-p)^{\Delta-m+1} {c}_m n.
\end{eqnarray*}

Taking into account that ${c}_{\Delta+1}=\pmatrix{\Delta \cr \Delta+1}=0$, we obtain
\begin{eqnarray*}
{\mathbf E}[|X|] &\ge& pn - \sum_{m=k}^{\Delta}(m-k+1) p^{m+1}
(1-p)^{\Delta-m} {c}_m n  
- \sum_{m=k+1}^{\Delta+1}(m-k)p^m (1-p)^{\Delta-m+1}
{c}_m n\\
&=& pn - \sum_{m=0}^{\Delta-k}(m+1) p^{m+k+1} (1-p)^{\Delta-m-k} {c}_{m+k} n\\
&& \qquad\qquad\qquad\qquad  - \sum_{m=0}^{\Delta-k}(m+1)p^{m+k+1} (1-p)^{\Delta-m-k}
{c}_{m+k+1} n\\
&=& 
pn - \sum_{m=0}^{\Delta-k}(m+1) p^{m+k+1} (1-p)^{\Delta-m-k}n \left({c}_{m+k} + {c}_{m+k+1}\right)\\
&=& 
pn - p^{k+1}n \sum_{m=0}^{\Delta-k}(m+1) {\tilde c}_{m+k+1} p^{m} (1-p)^{\Delta-k-m}.\\
\end{eqnarray*}

Furthermore, 
\begin{eqnarray*}
(m+1) {\tilde c}_{m+k+1} = {\Delta-k \choose m} {(m+1)! (\Delta+1)! \over (m+k+1)! (\Delta-k)! }\\
 \\
\le {\Delta-k \choose m} {(\Delta+1)! \over (k+1)! (\Delta-k)! } = {\Delta-k \choose m}{\tilde c}_{k+1}.
\end{eqnarray*}

We obtain
\begin{eqnarray*}
{\mathbf E}[|X|] &\ge&
pn - p^{k+1}n \sum_{m=0}^{\Delta-k} {\Delta-k \choose m} {\tilde c}_{k+1} p^{m} (1-p)^{\Delta-k-m}\\
&=& pn - p^{k+1}n {\tilde c}_{k+1} \\
&=& pn(1- p^{k} {\tilde c}_{k+1}) \\
&=& {k n \over {\tilde c}_{k+1} ^{1/k} \; (1+k)^{1+1/k}} \;.
\end{eqnarray*}
Since the expectation is an average value, there exists a particular $k$-limited packing of size at least 
${k n \over {\tilde c}_{k+1} ^{1/k} \; (1+k)^{1+1/k}}$, 
as required. The proof of the theorem is complete.
\end{pf}
\qed

The lower bound of Theorem \ref{th1} can be written in a simpler but weaker form as follows:

\begin{cor}
For any graph $G$ of order $n$,
$$
L_k(G) > {kn \over e (1+\Delta)^{1+1/k}}.
$$
\end{cor}
\begin{pf}
It is not difficult to see that
$$
{\tilde c}_{k+1} \le {(\Delta+1)^{k+1} \over (k+1)!}
$$
and, using Stirling's formula, 
$$
(k!)^{1/k} > \left(\sqrt{2\pi k}\left({k\over e}\right)^k\right)^{1/k} = \sqrt[2k]{2\pi k}\; {k\over e}.
$$
By Theorem \ref{th1},
$$
L_k(G) \ge {k n \left((k+1)!\right)^{1/k}  \over (\Delta +1)^{1+1/k} \; (1+k)^{1+1/k}} > {kn \over e (1+\Delta)^{1+1/k}} 
\times 
{\sqrt[2k]{2\pi k}\; k \over 1+k} >
{k n \over e (1+\Delta)^{1+1/k}}.
$$
Note that ${\sqrt[2k]{2\pi k}\; k \over 1+k} = {\sqrt[2k]{2\pi k} \over 1+1/k} > 1$. The last inequality is obviously true for $k=1$, while for $k\ge 2$ it can be rewritten in the equivalent form: $2\pi k > (1+1/k)^{2k} = e^2-o(1)$.
\end{pf}
\qed

\section{Randomized algorithm}

A pseudocode presented in Algorithm \ref{alg:k-limited} explicitly describes a randomized algorithm to find a $k$-limited packing  set, whose size satisfies bound (\ref{main_bound}) with a positive probability. Notice that Algorithm \ref{alg:k-limited} constructs a $k$-limited packing by recursively removing unwanted vertices from the initially constructed set $A$. This is different from the probabilistic construction used in the proof of Theorem 1. The recursive removal of vertices from the set $A$ may be more effective and efficient, especially if one tries to remove overall as few vertices as possible from $A$ by maximizing intersections of the sets $A'_m$ ($m=k,\ldots,\Delta$) and $B'_m$ ($m=k+1,\ldots,\Delta$). 


Algorithm \ref{alg:k-limited} can be implemented to run in $O(n^2)$ time. To compute the probability $\displaystyle{p={1 \over \sqrt[k]{\pmatrix{\Delta \cr k} (\Delta +1)}}}$, the binomial coefficient $\pmatrix{\Delta \cr k}$ can be computed by using the dynamic programming and Pascal's triangle in $O(k\Delta)=O(\Delta^2)$ time using $O(k)=O(\Delta)$ memory. The maximum vertex degree $\Delta$ of $G$ can be computed in $O(m)$ time, where $m$ is the number of edges in $G$.
Then $p$ can be computed in $O(m+\Delta^2)=O(n^2)$ steps. It takes $O(n)$ time to find the initial set $A$. Computing the intersection numbers $r=|N(v)\cap A|$ and removing unwanted vertices of $N(v)\cap A$'s from $A$ can be done in $O(n+m)$ steps. Finally, checking whether $X$ is maximal or extending $X$ to a maximal $k$-limited packing can be done in $O(n+m)$ time: try to add vertices of $V(G)-A$ to $X$ recursively one by one, and check whether the addition of a new vertex $v\in V(G)-A$ to $X$ violates the conditions of a $k$-limited packing for $v$ or at least one of its neighbours in $G$ with respect to $X\cup\{v\}$. Thus, overall Algorithm \ref{alg:k-limited} takes $O(n^2)$ time, and, since $m=O(n^2)$ in general, it is linear in the graph size ($m+n$) when $m=\theta(n^2)$.


\begin{algorithm}\label{alg:k-limited}
    \caption{Randomized $k$-limited packing}

    \KwIn{Graph $G$ and integer $k$, $1\le k\le
\Delta$.}
    \KwOut{$k$-Limited packing $X$ in $G$.}
    \BlankLine
\Begin{

Compute $p = \left({1 \over {\tilde c}_{k+1} \; (1+k)}\right)^{1/k}\ $\;
\SetLine
Initialize $A=\emptyset$;\tcc*[f]{Form a set $A\subseteq V(G)$}\\
\ForEach{ vertex $v\in V(G)$} {
    with the probability $p$, decide whether $v\in A$ or
$v\notin A$\; }
\SetLine
\tcc*[f]{Recursively remove redundant vertices from $A$}\\
\ForEach{ vertex $v\in V(G)$} {
 Compute $r=|N(v)\cap A|$\;
    \If{$v\in A$ and $r\ge k$}
        {remove any $r-k+1$ vertices of $N(v)\cap A$ from $A$\;
    }
    \If{$v\notin A$ and $r>k$}
    	{remove any $r-k$ vertices of $N(v)\cap A$ from $A$\;
    }
}
Put $X = A$;\tcc*[f]{$A$ is a $k$-limited packing}\\
Extend $X$ to a maximal $k$-limited packing\;
\Return $X$;
}
\end{algorithm}

Also, this randomized algorithm for finding $k$-limited packings in a graph $G$ can be implemented in parallel or as a local distributed algorithm. As explained in \cite{GPZ13}, this kind of algorithms are especially important, e.g. in the context of ad hoc and wireless sensor networks.
We hope that this approach can be also extended to design self-stabilizing or on-line algorithms for $k$-limited packings. For example, a self-stabilizing algorithm searching for maximal $2$-packings in a distributed network system is presented in \cite{S12}. Notice that self-stabilizing algorithms are distributed and fault-tolerant, and use the fact that each node has only a local view/knowledge of the distributed network system. This provides another motivation for efficient distributed search and algorithms to find $k$-limited packings in graphs and networks.

\section{Sharpness of the bound of Theorem \ref{th1}}

We now show that the lower bound of Theorem \ref{th1} is asymptotically best possible for some values of $k$. Let $\delta$ = $\delta(G)$ denote the minimum vertex degree in a graph $G$. The bound of Theorem \ref{th1} can be rewritten in the following form for $\Delta \ge k$:
$$
L_k(G) \ge {k n \over (k+1) \sqrt[k]{\pmatrix{\Delta \cr k} (\Delta +1)} }.
$$
Combining this bound with the upper bound of Lemma $8$ from \cite{GGHR10}, we obtain that 
for any connected graph $G$ of order $n$ with $\delta(G)\ge k$,
\begin{equation} \label{ineq5}
{1 \over \sqrt[k]{\pmatrix{\Delta \cr k} (\Delta +1)} }\times {k \over k+1} n  \;\le\;
L_k(G)\; \le \;{k \over k+1} n.
\end{equation}
Notice that the upper bound in the inequality (\ref{ineq5}) is sharp (see \cite{GGHR10}), so these bounds  provide an interval of values for $L_k(G)$ in terms of $k$ and $\Delta$ when $k\le\delta$. For regular graphs, $\delta=\Delta$, and, when $k=\Delta$, we have
$$
{1 \over \sqrt[k]{\pmatrix{\Delta \cr k} (\Delta +1)} } = 
{1 \over \left(k+1\right)^{1/k}} \longrightarrow 1\ \quad\mbox{as} \quad \ k \to \infty.
$$
Therefore, Theorem \ref{th1} is asymptotically sharp in regular connected graphs for $k=\Delta$. 
A similar statement can be proved for the situation when $k=\Delta (1-o(1))$. 

Thus, the following result is true:

\begin{thm} \label{asym_sharp}
When $n$ is large there exist graphs such that 
\begin{equation} 
L_k(G) \le {k n \over {\tilde c}_{k+1}
^{1/k} \; (1+k)^{1+1/k}}(1+o(1)).
\end{equation}
\end{thm}
 

\section{Upper bounds}

As mentioned earlier, $\rho(G)=L_1(G)\le\gamma(G)$. In \cite{GGHR10}, the authors provide several upper bounds for $L_k(G)$, e.g. $L_k(G)\le k\gamma(G)$ for any graph $G$. Using the well-known bound (see e.g. \cite{AS00})
$$
\gamma(G) \le {\ln(\delta+1)+1 \over \delta+1} n,
$$
we obtain
\begin{equation}\label{classical}
L_k(G)\le {\ln(\delta+1)+1 \over \delta+1} kn.
\end{equation}
Even though this bound does not work well when $k$ is `close' to $\delta$, it is very reasonable for small values of $k$. 

We now prove an upper bound for the $k$-limited packing number in terms of the $k$-tuple domination number. 
A set $X$ is called a {\it
$k$-tuple dominating set} of $G$ if for every vertex $v\in V(G)$,
$|N[v]\cap X|\ge k$. The minimum cardinality of a $k$-tuple
dominating set of $G$ is the {\it $k$-tuple domination number}
$\gamma_{\times k}(G)$. The $k$-tuple domination number is only
defined for graphs with $\delta\ge k-1$. 

\begin{thm} \label{th2}
For any graph $G$ of order $n$ with $\delta\ge k-1$,
\begin{equation} \label{upper_bound}
L_k(G)\le\gamma_{\times k}(G).
\end{equation}
\end{thm}

\begin{pf}
We prove inequality (\ref{upper_bound}) by contradiction.
Let $X$ be a maximum $k$-limited packing in $G$ of size $L_k(G)$, and
let $Y$ be a minimum $k$-tuple dominating set in $G$ of size $\gamma_{\times k}(G)$. 
We denote $B=X\cap Y$, i.e. $X=A\cup B$ and $Y=B\cup C$.
Assume to the contrary that $L_k(G)>\gamma_{\times k}(G)$, thus $|A|>|C|$.

Since $Y$ is $k$-tuple dominating set, each vertex of $A$ is adjacent to at least $k$ vertices of $Y$. Hence the number of edges between $A$ and $B\cup C$ is as follows: 
$$e(A,B\cup C) \ge k|A|.$$
Now, every vertex of $C$ is adjacent to at most $k$ vertices of $X$, because $X$ is a $k$-limited packing set. Therefore, the number of edges between $C$ and $A\cup B$ satisfies
$$e(C,A\cup B) \le k|C|.$$
We obtain
$$
e(C,A\cup B) \le k|C| < k|A| \le e(A,B\cup C),
$$
i.e. $e(C,A\cup B) < e(A,B\cup C)$. By eliminating the edges between $A$ and $C$, we conclude that
$$
e(C,B) < e(A,B).
$$

Now, let us consider an arbitrary vertex $b\in B$ and denote $s=|N(b)\cap A|$. Since $X=A\cup B$ is a $k$-limited packing set, we obtain $|N(b)\cap X|\le k-1$, and hence $|N(b)\cap B|\le k-s-1$. On the other hand, $Y=B\cup C$ is $k$-tuple dominating set, so $|N(b)\cap Y|\ge k-1$. Therefore, $|N(b)\cap C|\ge s$. Thus, $|N(b)\cap C|\ge |N(b)\cap A|$ for any vertex $b\in B$. We obtain,
$$
e(C,B) \ge e(A,B),
$$
a contradiction. We conclude that  $L_k(G)\le \gamma_{\times k}(G)$.
\end{pf}
\qed

Notice that it is possible to have $k=\Delta+1$ in the statement of Theorem \ref{th2}, which is not covered by Theorem \ref{th1}. Then $\delta=\Delta$, which implies the graph is regular. However, $L_k(G)=\gamma_{\times k}(G)=n$ for $k=\delta+1=\Delta+1$. In non-regular graphs, $\delta+1\le\Delta$, and $k\le\Delta$ to satisfy the conditions of Theorem \ref{th1} as well.

For $t\le \delta$, we define
$$
\delta' = \delta -k +1 
 \quad \mbox{and} \quad
 {\tilde b}_t = {\tilde b}_t (G) =  \pmatrix{\delta+1 \cr t}.
$$
Using the upper bound for the $k$-tuple domination number from \cite{GPZ13}, we obtain:

\begin{cor} \label{cor2}
For any graph $G$ with $\delta\ge k$,
\begin{equation}
L_k(G)\le  \left(1-{\delta' \over {\tilde b}_{k-1}
^{1/\delta'} (1+\delta')^{1+1/\delta'}}\right) n.
\end{equation}
\end{cor}

In some cases, Theorem \ref{th1} and Corollary \ref{cor2} simultaneously provide good bounds for the $k$-limited packing number. For example, for a 40-regular graph $G$:
$$
0.312n < L_{25}(G)  < 0.843n.
$$

\end{document}